\documentclass[aps,pre, showpacs, showkeys, twocolumn, superscriptaddress]{revtex4-1}

\usepackage{amssymb, amsmath, amsfonts, amsthm, mathtools}
\usepackage{graphicx, xcolor}
\usepackage{empheq}
\usepackage[T1]{fontenc}
\usepackage{xspace}

\DeclareMathOperator{\erf}{erf}

\renewcommand*{\d}{{\mathrm d}}

\newcommand*{\vbar}{\overline{v} \xspace}

\newcommand{\eq}[1]{Eq.~#1}
\newcommand{\eqs}[1]{Eqs.~#1}
\newcommand{\fig}[1]{Fig.~#1}

\newcommand{\Sec}[1]{Sec.~#1}

\begin{document}


\title{Sign changes as a universal concept in first-passage time calculations}


\author{Wilhelm Braun}
\email{wilhelm.braun@uottawa.ca}
\affiliation{Department of Physics, University of Ottawa, Ottawa, Canada K1N 6N5}

\author{R\"{u}diger Thul}
\affiliation{Centre for Mathematical Medicine and Biology, School of Mathematical Sciences, University of Nottingham, Nottingham, NG7 2RD, UK}


\date{\today}

\begin{abstract}

First-passage time problems are ubiquitous across many fields of study including transport processes in semiconductors and biological synapses, evolutionary game theory and percolation. Despite their prominence, first-passage time calculations have proven to be particularly challenging.  Analytical results to date have often been obtained under strong conditions, leaving most of the exploration of first-passage time problems to direct numerical computations. Here we present an analytical approach that allows the derivation of first-passage time distributions for the wide class of non-differentiable Gaussian processes. We demonstrate that the concept of sign changes naturally generalises the common practice of counting crossings to determine first-passage events. Our method works across a wide range of time-dependent boundaries and noise strengths thus alleviating common hurdles in first-passage time calculations.

\end{abstract}


\maketitle


\section{Introduction}
A frequent question in the study of stochastic systems is to determine when a particular event happens for the first time. Such first-passage time (FPT) problems are ubiquitous in the physical sciences, and applications include trapping reactions \cite{Bray:2007}, earthquakes \cite{Matthews:2002}, goodness-of-fit tests \cite{Chicheportiche:2012}, habitat selection \cite{Fauchald:2003}, dark matter halos \cite{Maggiore:2010}, decision making \cite{Srivastava:2015, marshall_et_al_2009}, diffusion in complex environments \cite{condamin_benichou_moreau_2005, guerin_et_al_2016} and ion channel dynamics \cite{gerstein_mandelbrot_1964,Goychuk:2002,Thul:2007gy}. For a recent review on the wider applications of FPTs see \cite{Bray:2013ha,FPT:2014}. Although the concept of FPTs is easily stated, the actual computation poses substantial challenges. Researchers can now draw on an extended suite of mathematical techniques to evaluate FPTs, ranging from analytical descriptions for special cases including asymptotic expansions to semi-analytical approaches based on integral equations to direct numerical simulations \cite{siegert_1951,bressloff_book_stochastic_processes,CKbook,redner_book,alili_patie_pedersen_2005,Atiya:2005,lindner_et_al_2003,urdapilleta_pre_2011,touboul_faugeras_2008, gs_review,tuckwell_1, ttmm_2013,evans_majumdar_2011,reuveni_2016,newby_allard_2016,Holcman:2014ip, chacron_lindner_longtin_2004}. Notwithstanding these practical methods, the general solution to one of the most fundamental FPT problems is still unknown: diffusion to an arbitrary boundary \cite{gs_review,Bujorianu:2012,nobile_pirozzi_ricciardi_2007,Tamborrino:2016db}. This is even more remarkable since diffusion processes are often used to approximate the dynamics of more complicated stochastic systems \cite{gardiner,bressloff_book_stochastic_processes}.

The seminal work by Wiener and Rice \cite{rice, cramer_leadbetter_book} has been instrumental in advancing our understanding of FPT problems. They derived a series representation of the FPT distribution for a sufficiently smooth stationary stochastic process through a constant boundary.  A common interpretation of this smoothness condition is that the stochastic process is differentiable in the mean-square sense.  For a stationary stochastic process, differentiability  arises from the behaviour of its covariance function at the origin. If it is differentiable there, the process is differentiable in the mean square sense \cite{Papoulis:2006}. Interestingly, this condition is already violated for any diffusion process based on Brownian motion \cite{gs_sde_a}, most notably the ubiquitous Ornstein-Uhlenbeck process (OUP) and geometric Brownian motion.

A distinct advantage of the Wiener--Rice approach is that it often gives rise to compact analytical expressions, which in turn provide great insight into the stochastic system under investigation, as e.g. illustrated in \cite{verechtchaguina_2006,badel_pre_2011}. In this paper, we show how the concepts that underlie the Wiener--Rice series can be generalised to non-differentiable stochastic processes with arbitrary boundaries, thus translating the versatility and efficacy of the Wiener--Rice approach to a much enhanced range of physically relevant dynamics.

\section{Sign changing probabilities}
\label{sec:sign}

Consider a stationary stochastic process $X(t)$ and a time-dependent boundary $S(t)$. Let $N([0,t])$ denote the number of crossings of $X(t)$ and $S(t)$ in a fixed interval $[0,t]$. If $X(t)$ is differentiable, the mean number of crossings of $X(t)$ through $S(t)$ in any given finite time interval is always finite, i.e. $\mathbb E\{N([0,t])\}<\infty$ \cite{rice,cramer_leadbetter_book}. This entails that we can count the number of crossings, and in particular that there is at most one crossing within a sufficiently small time interval of length $\Delta$. When we introduce the new process $Z(t)=X(t)-S(t)$, we see that $Z(t)$ changes sign as soon as $X(t)$ crosses through $S(t)$, and hence instead of counting crossings, we could count sign changes. If $X(t)$ is non-differentiable, Rice's seminal work shows that  $\mathbb E\{N([0,t])\}=\infty$. While this seems counterintuitive, infinite means are not uncommon in physics. A prominent example are power-law probability distributions of the form $p(x) \sim x^{-\alpha}$, $\alpha>0$ \cite{Newman:2005gv} \footnote{For $\alpha \leq 2$, the mean of the probability distribution diverges, yet the distribution of the intensity of solar flares or the distribution of family names are well captured by power laws with $\alpha \leq 2$ \cite{Lu:1991,Zanette20011}}.
The divergence of $\mathbb E\{N([0,t])\}$ does not allow us to count the number of crossings or sign changes anymore. In particular, the divergence holds for any $t >0$, so we cannot choose an infinitesimal time interval $\Delta$ to overcome it. In turn, this prevents us from using the original results by Wiener and Rice. 

Progress can be made here by a change of perspective. We generally identify a crossing event in some interval $[t,t+\Delta]$ with $Z(t)$ and $Z(t+\Delta)$ having opposite signs, i.e. we evaluate $Z$ at the \emph{end points} of the interval. For sufficiently small $\Delta$ and a differentiable stochastic process, this is equivalent of a crossing anytime \emph{during} the interval and hence justifies the practice of evaluating $Z$ at the end points. For a non-differentiable stochastic process, we can still evaluate $Z(t)$ and $Z(t+\Delta)$ and check for sign changes. The only difference is that this does not inform us about the number of crossings, but about the existence of \emph{at least} one crossing. Given a sufficiently small $\Delta$, this is the key information for practical calculations. Sign changes of $Z(t)$ and $Z(t+\Delta)$ can therefore be used to identify FPTs for both differentiable and non-differentiable stochastic processes.

\begin{figure}[htb]
\centering
\includegraphics[width = 0.45\textwidth]{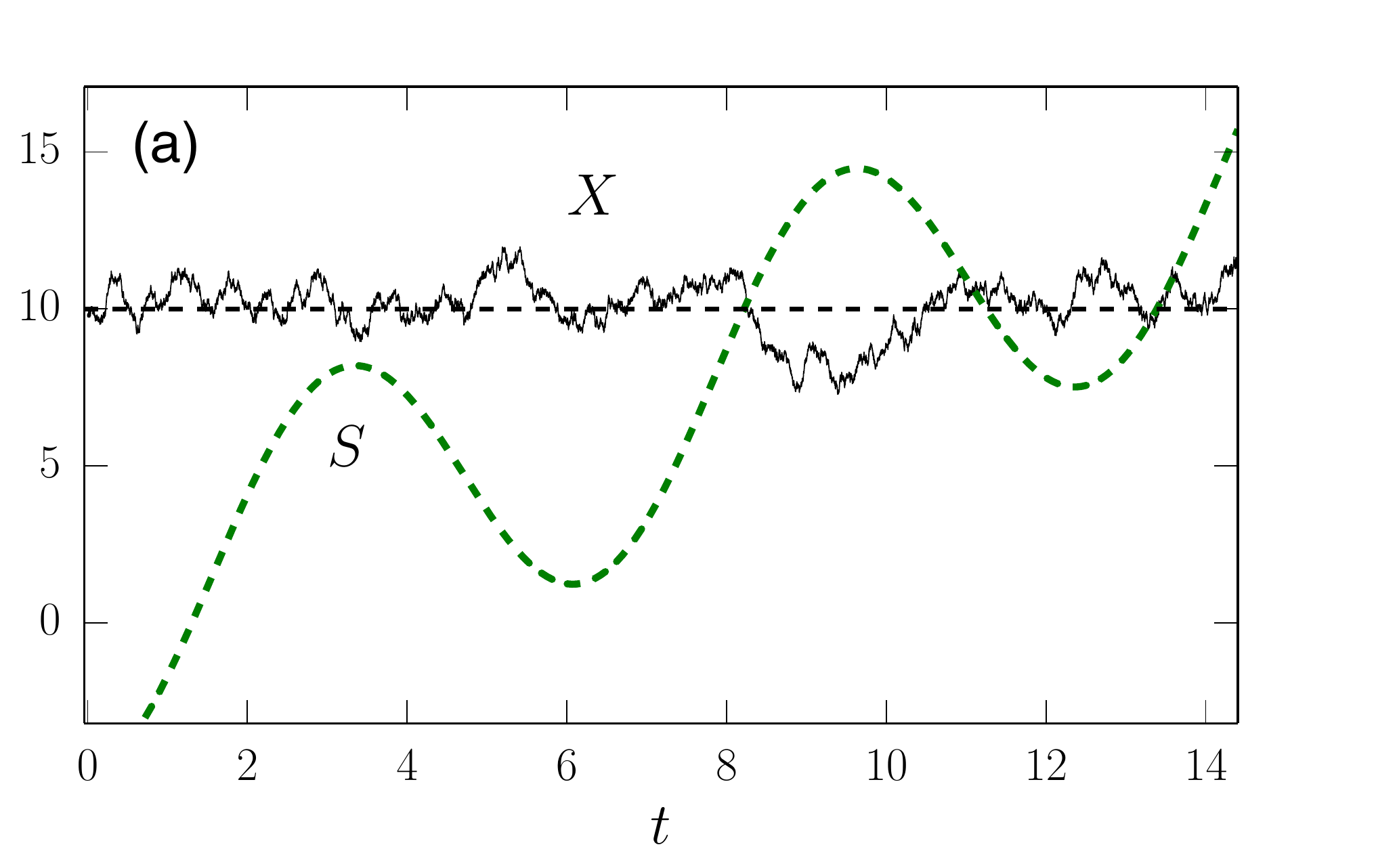}
\includegraphics[width = 0.45\textwidth]{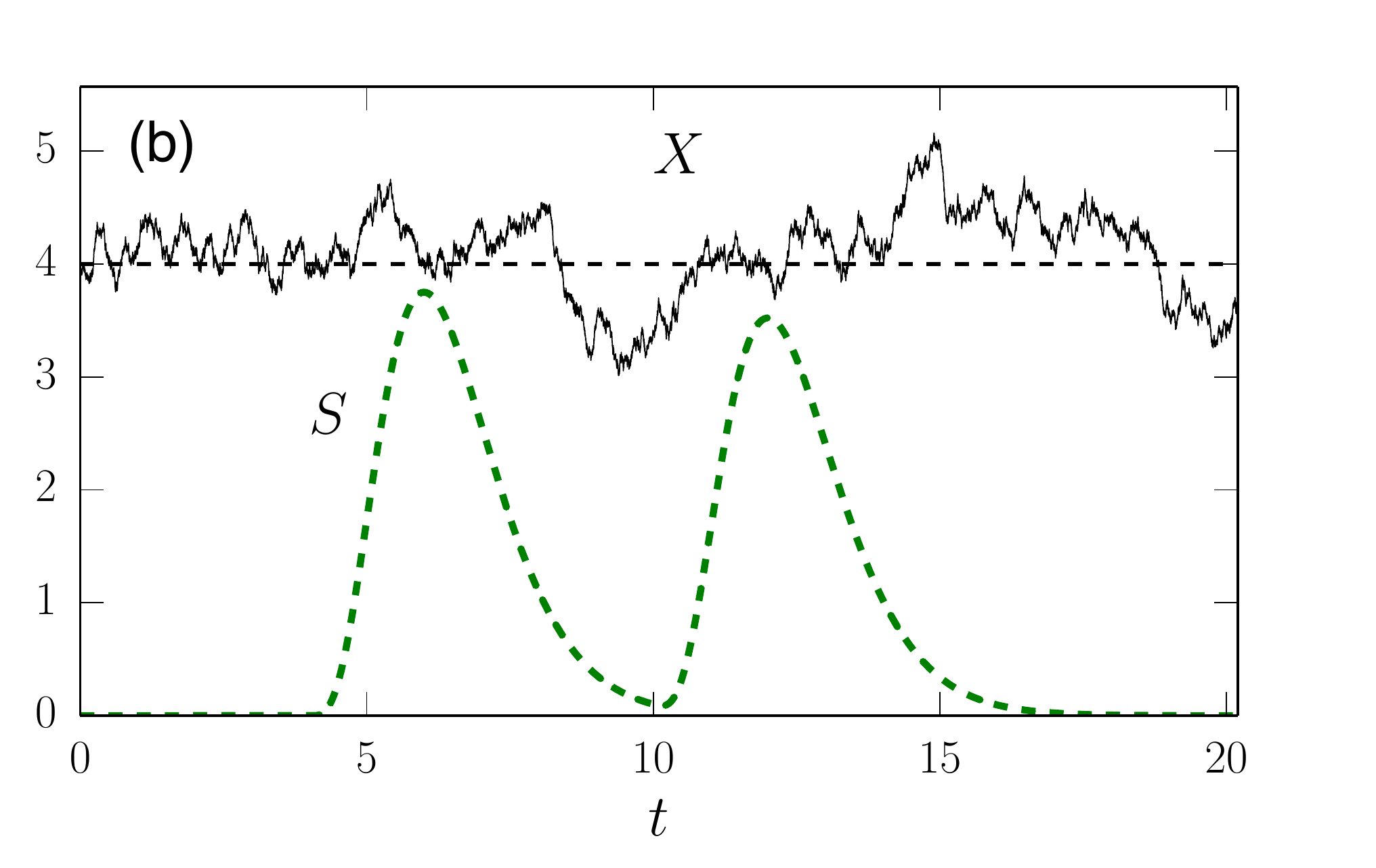}
\caption{An OUP (solid black line) with mean $\vbar$ (dashed black line) crosses through time dependent boundaries $S(t)$ (green dashed line). The boundaries are (a) $S(t)= -5\cos(t)\exp\left(-(0.1\cos(t))^{2} \right) +  t$ and (b) $S(t)=\sum_n \alpha_n (2(t-T_n))^4 \exp(-2(t-T_n)) H(t-T_n)$ with $\alpha \in \{0.8,0.75\}$, $T \in \{4,10\}$. Parameter values are (a) $\tau = 1.0$, $\vbar = 10.0$, $D=2.0$ and (b) $\tau = 2.0$, $\vbar = 4.0$, $D=0.2$. $H$ denotes the Heaviside step function.}
\label{fig:ensemble_sketch}
\end{figure}

The versatility of sign changes in computing FPTs is best illustrated with cases that cannot be solved with existing analytical techniques, yet frequently occur in applications. Figure \ref{fig:ensemble_sketch} shows prototypical examples to which current approaches cannot be applied since the boundaries are neither convex nor concave. These boundaries can be found in areas as diverse as neuronal dynamics (\fig \ref{fig:ensemble_sketch}a) and molecular transition theory (\fig \ref{fig:ensemble_sketch}b). It is worth pointing out that the boundary in \fig \ref{fig:ensemble_sketch}a originates from transforming a FPT problem for a driven stochastic system with constant threshold. In general, these problems cannot be solved analytically due to the explicit time-dependence of the stochastic dynamics \cite{touboul_faugeras_2007}. However, by shifting the time-dependence to the threshold as we do here, the problem becomes analytically tractable. The example in \fig \ref{fig:ensemble_sketch}b illustrates motion in a complex energy landscape with a time-dependent metastable state \cite{Hanggi:1990en,Ebeling:2003ws}. While this is most reminiscent of problems in physical chemistry, the same scenario occurs in general phase space dynamics \cite{Wiggins:2003wd}.

The specific FPT problem that we will investigate is to determine when a Gaussian process $X(t)$, whose initial value is drawn from its stationary distribution, crosses through $S(t)$ from above for the first time, i.e. we are interested in the random variable $T = \inf\left\{t>0 |X(t) \leq S(t)\right\}$ with a corresponding probability distribution ${\cal F}(t)$. While the theory presented here is valid for any Gaussian process, we will illustrate our results with an OUP given its prominence across so many  fields of study. The dynamics of the OUP is governed by 
\begin{equation}
\d X=-\frac{X-\vbar}{\tau}d t+\sqrt D d W(t)\,,
\end{equation}
where $W(t)$ is a standard Brownian motion, $\vbar$ is the mean of the OUP and $\tau$, $D>0$. The stationary variance and correlation coefficient of the OUP are given by $D \tau/2$ and $\exp(-|t|/\tau)$, respectively.  

\begin{figure}
\centering
\includegraphics[width = 0.45\textwidth]{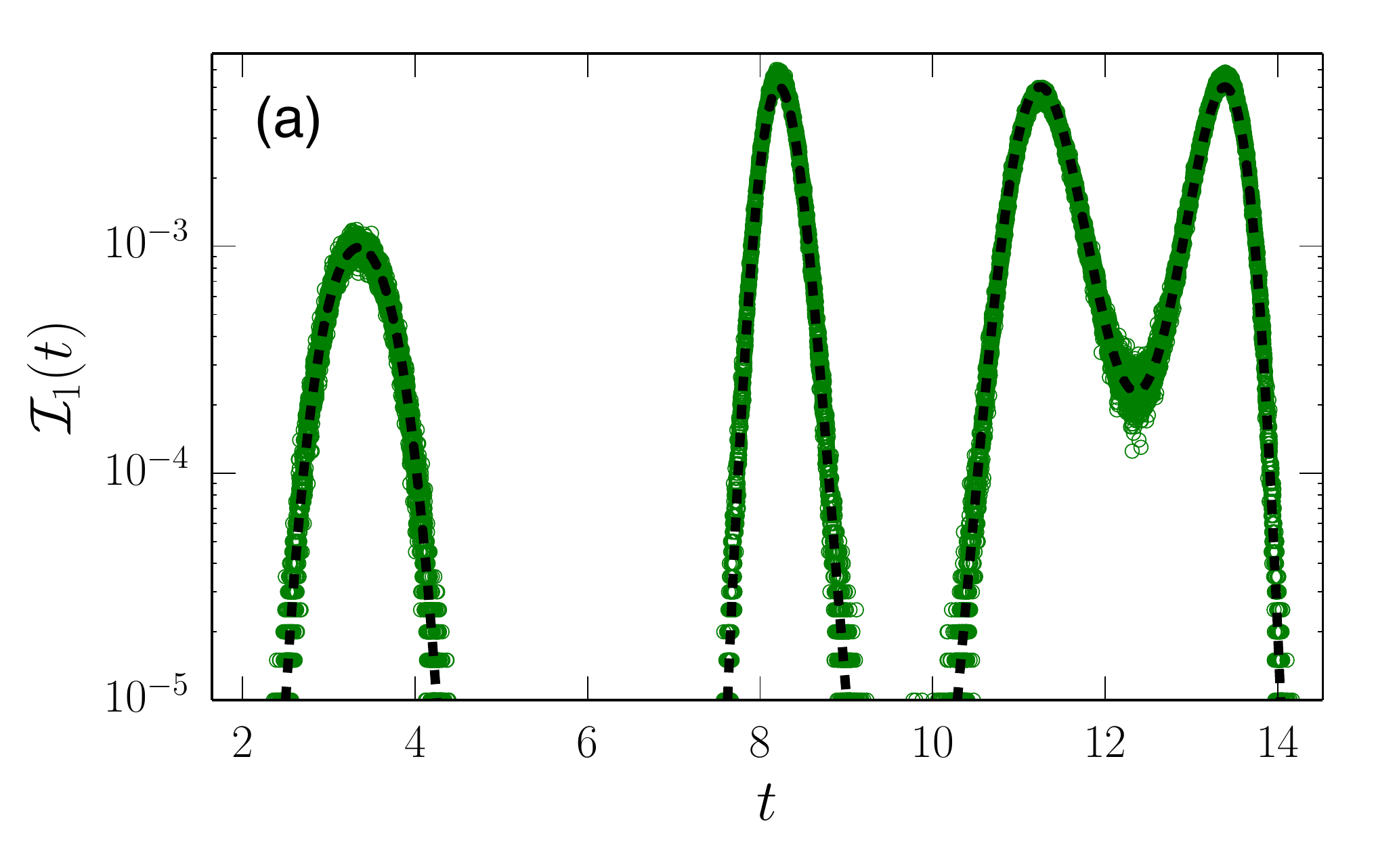}
\includegraphics[width = 0.45\textwidth]{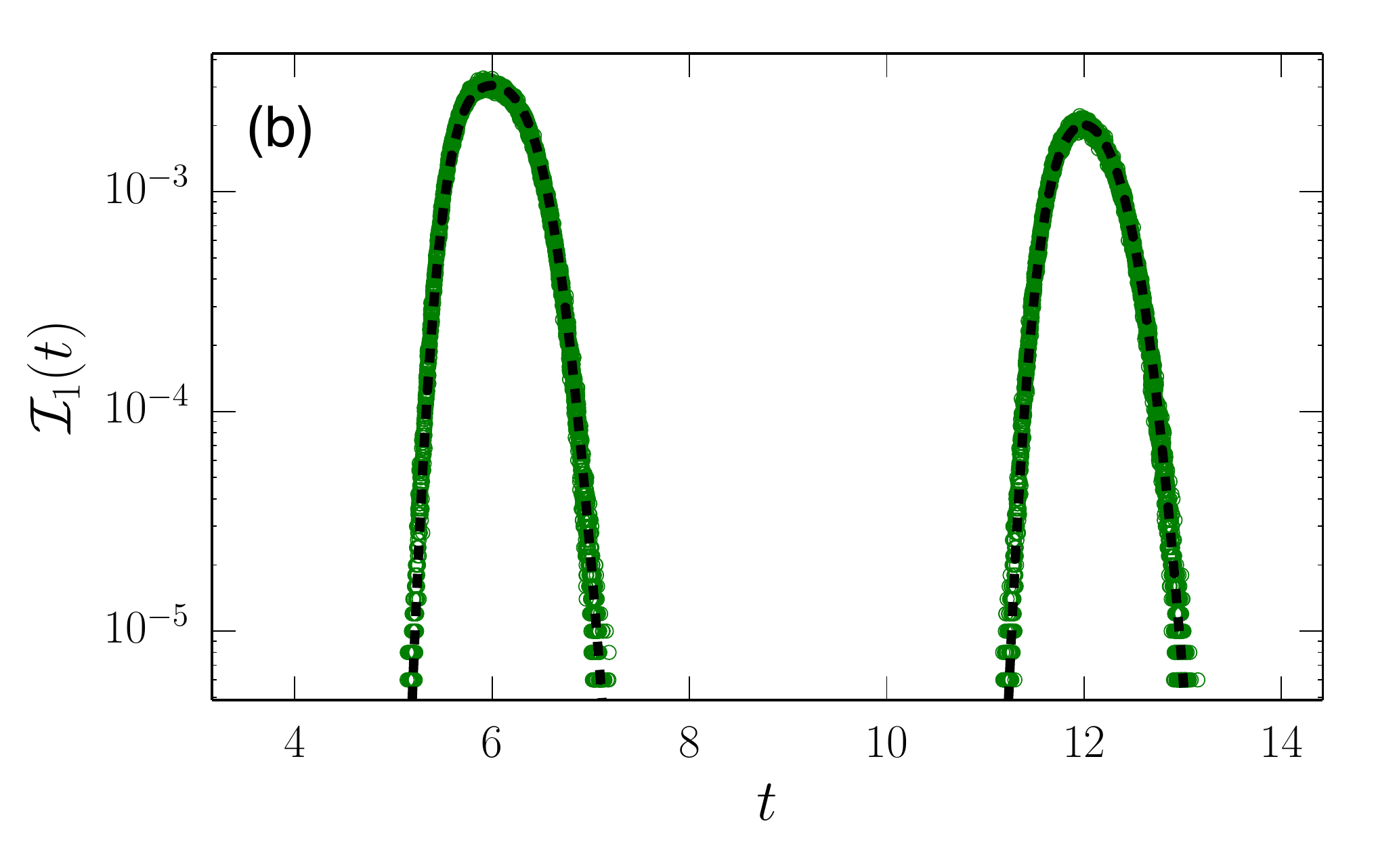}
\caption{Crossing probability ${\cal I}_1(t)$ for the two cases shown in \fig \ref{fig:ensemble_sketch} obtained from MC simulations (green circles) and from \eq \eqref{eq:I_1} (black dashed line).  
}
\label{fig:rate_comparison}
\end{figure}

To begin our analysis, we define the conditional sign change probability $p_{+-}(t| x_0)= \mathbb P(Z(t)>0,Z(t+\Delta)<0|x_0)$, i.e. the probability of a sign change in $[t,t+\Delta]$ given a value of  $X(0)=x_0$. 
%
Let $f_2(x_1,x_2|x_0)=\mathbb P(X(t_1)=x_1, X(t_2)=x_2 |X(0)=x_0)$ denote the conditional bivariate probability function of $X(t)$, we see that
\begin{equation}
\begin{split}
p_{+-}(t|x_0)
=\int_{S(t)}^{\infty} \!\!\! d x_{1}\int_{-\infty}^{S(t+\Delta)}d x_{2} \, f_2(x_{1},x_{2}|x_{0}) \,.
\end{split}
\label{eq:pmp_td}
\end{equation}
Note that \eq \eqref{eq:pmp_td} is always well defined and does not rely on whether $X(t)$ is differentiable or not. Since $X(t)$ is Gaussian, $f_2$ is Gaussian again, and the corresponding mean and covariance matrix can be obtained in closed form. By expanding the integrals in \eq \eqref{eq:pmp_td} to lowest orders in $\Delta$ following \cite{mcfadden_1967}, we find 
\begin{equation}
p_{+-}(t|x_0)=
\frac 1 \pi \sqrt{\! \frac{\Delta/\tau}{2[1-\rho(t)^2]}} \exp\!\left [\!\frac{[S(t)-m_1(t)]^2}{2 \sigma_1^2(t)}\! \right],
\label{eq:p_pm_gauss}
\end{equation}
where $m_1(t)=\vbar+(x_0-\vbar)\rho(t)$, $\sigma_1^2(t)=\sigma^2(1-\rho(t)^2)$, and $\rho(t)$ and $\sigma^2$ denote the correlation coefficient and the variance of $X(t)$, respectively. Hence, $p_{+-}(t|x_0)=\sigma\sqrt{\Delta/(\pi \tau)}f_1(S(t)|x_0)$, where $f_1$ presents a conditional univariate Gaussian probability function. The value of $x_0$ has been fixed but arbitrary so far. By integrating it out we obtain the probability ${\cal I}_1(t)=\int_0^\infty p_{+-}(t|x) p(x) d x$  for a sign change in the interval $[t,t+\Delta]$. Since both $p(x)$ and $p_{+-}(t|x)$ are Gaussian, the integral can be performed analytically, and we arrive at
\begin{equation}
\mathcal I_{1}(t)=
\sqrt{\frac{\Delta}{8\pi^2 \tau}} \exp[e(t)] (1+\erf[f(t)])\,,
\label{eq:I_1}
\end{equation}
where $2 e(t)=\omega(t)^2 \sigma_1^2(t)-(S(t)-\vbar)^2/\sigma_1^2(t)$, $f(t)=(\vbar-\omega(t) \sigma_1^2(t))/\sqrt{2 \sigma_1^2(t)}$ and $\omega(t)=\rho(t)(\vbar-S(t))/\sigma_1^2(t)$.
We compare the analytical expression for ${\cal I}_1$ with direct Monte-Carlo (MC) simulations in \fig \ref{fig:rate_comparison} for the two choices of $S(t)$ shown in \fig \ref{fig:ensemble_sketch}. The agreement is excellent, capturing the multimodal character of the crossing probability induced by the non-monotonicity of $S(t)$  and spanning more than three orders of magnitude, with probabilities reaching values smaller than $10^{-5}$.

\section{FPT probabilities}

\begin{figure}[t]
\centering
\includegraphics[width = 0.225\textwidth]{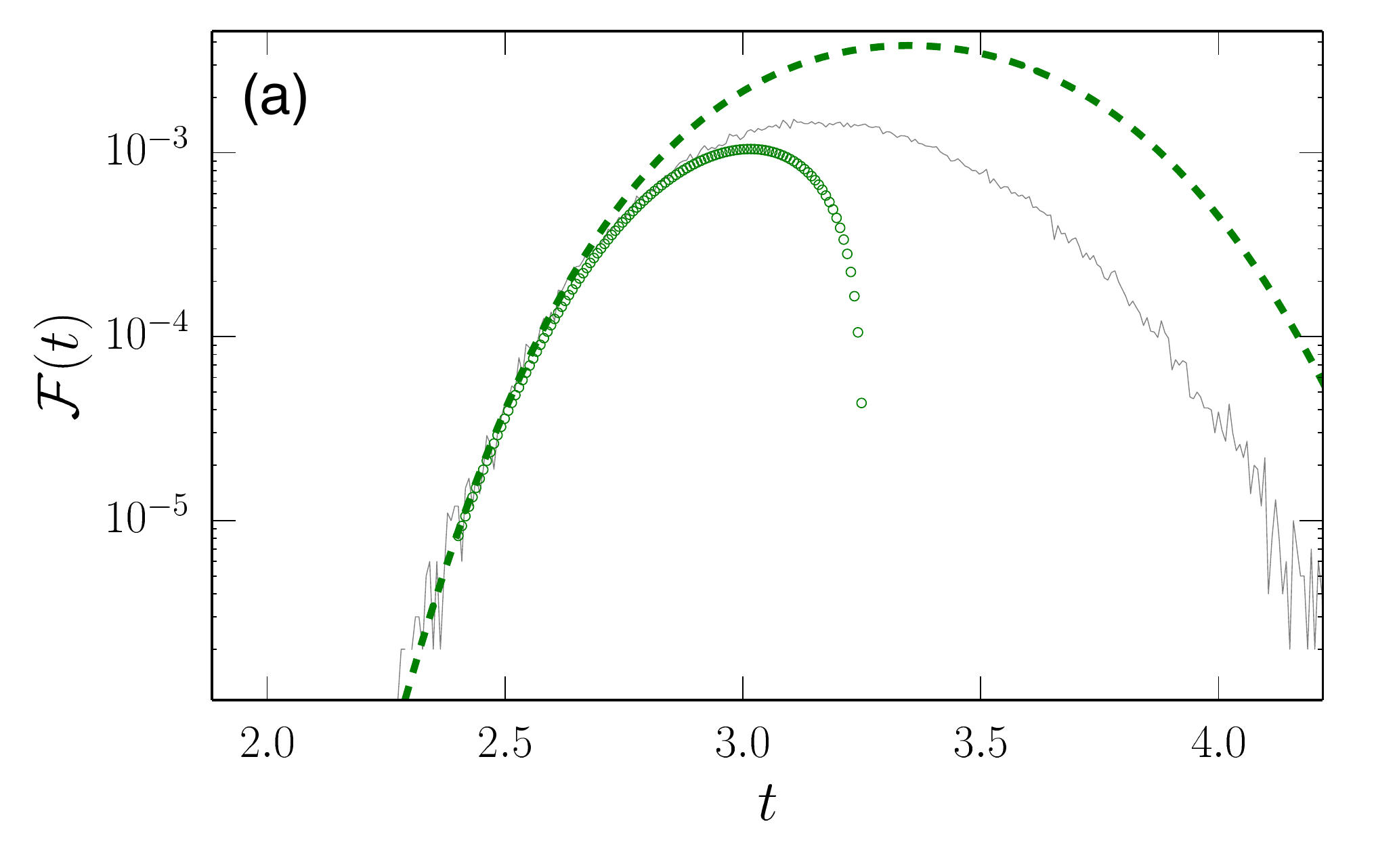}
\includegraphics[width = 0.225\textwidth]{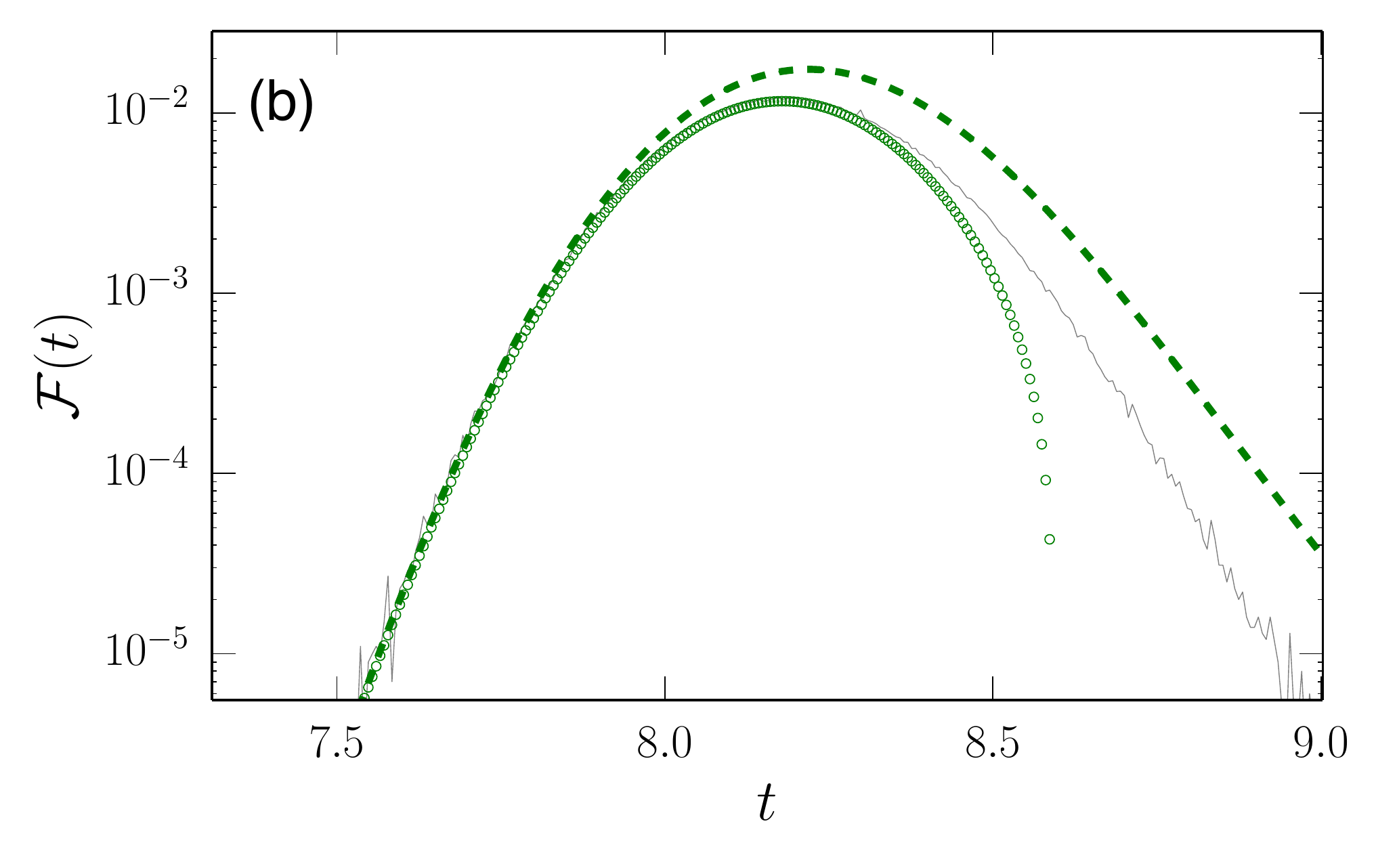}\\
\includegraphics[width = 0.225\textwidth]{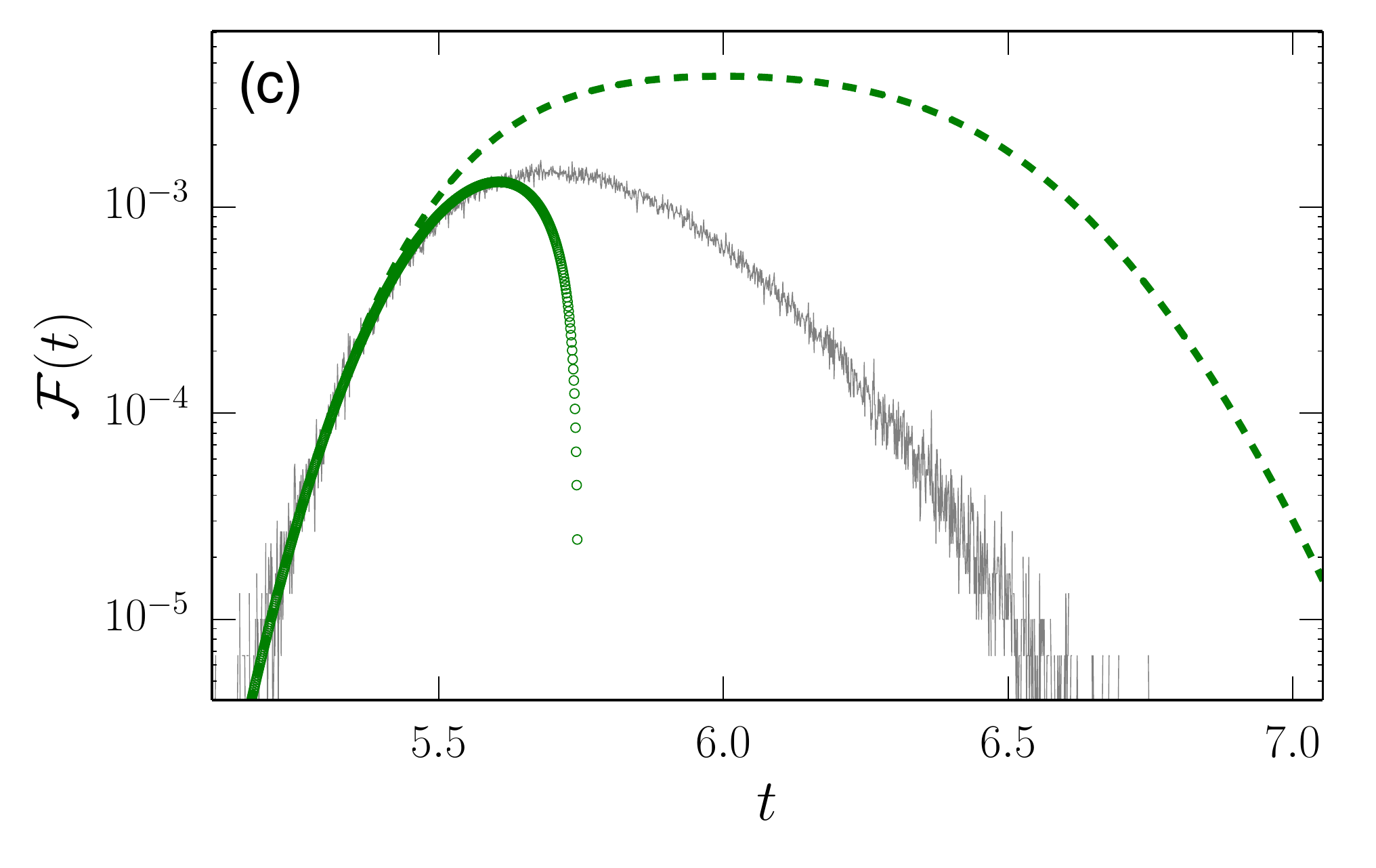}
\includegraphics[width = 0.225\textwidth]{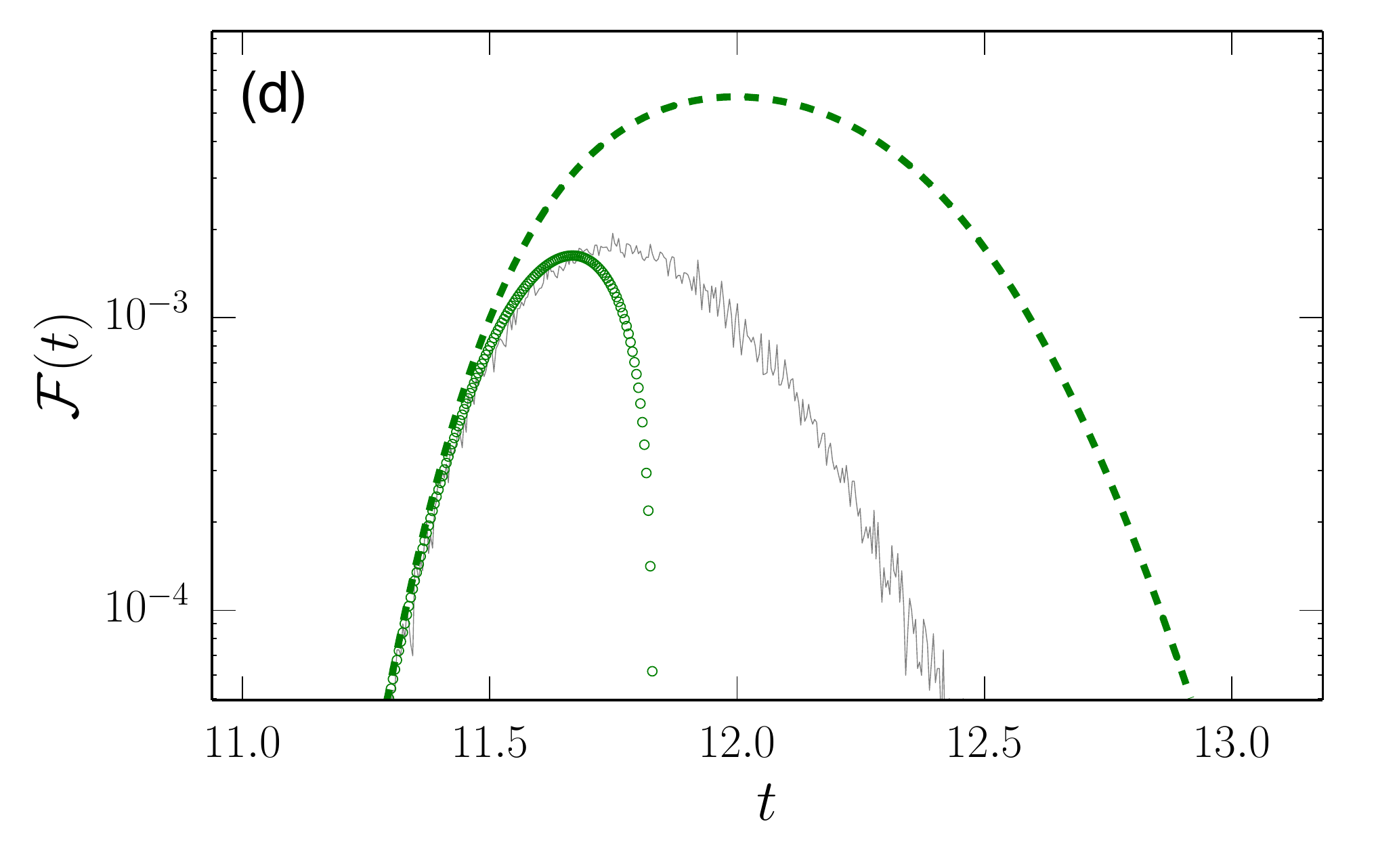}
\caption{FPT probability $\mathcal F(t)$ corresponding to \fig \ref{fig:ensemble_sketch}a (a,b) and \fig \ref{fig:ensemble_sketch}b (c,d) obtained from MC simulations (solid grey line), \eq \ref{eq:I_1} (dashed green line) and \eq \ref{eq:G_2} (green circles). 
}
\label{fig:fpt_probabilities}
\end{figure}

From a conceptual point of view, ${\cal I}_1(t)$ corresponds to the first term ${\cal J}_1(t)$ of the FPT expansion derived by Wiener and Rice. Their original results for the full FPT distribution ${\cal F}(t)$ that measures crossings in the interval $[t,t+\Delta]$ can be expressed as 
\begin{equation}
{\cal F}(t)=\sum_{n=0}^\infty \frac{(-1)^n}{n!} {\cal J}_{n+1}(t)\,,
\end{equation}
where ${\cal J}_n(t)=\sum_{k=n}^\infty q_k(t) (k-1)!/(k-n)!$ and $q_n(t)$ denotes the probability of $n$ crossings during the time $[0,t+\Delta]$ including one in $[t,t+\Delta]$. Therefore, ${\cal J}_1(t)=\sum_{k=1}^\infty q_{k}(t)$  represents the probability of a crossing in $[t,t+\Delta]$ irrespective of how many other crossings have occurred prior to $t$. In comparison, ${\cal I}_1(t)$ measures the probability of a sign change in $[t,t+\Delta]$ regardless of the behaviour of $X(t)$ before $t$. Therefore, we expect ${\cal I}_1(t)$ to constitute a first order approximation ${\cal G}_1(t)$ of ${\cal F}(t)$: ${\cal F}(t) \approx {\cal G}_1(t)={\cal I}_1(t)$. This is confirmed in \fig \ref{fig:fpt_probabilities}.

\begin{figure}
\centering
\includegraphics[width = 0.45\textwidth]{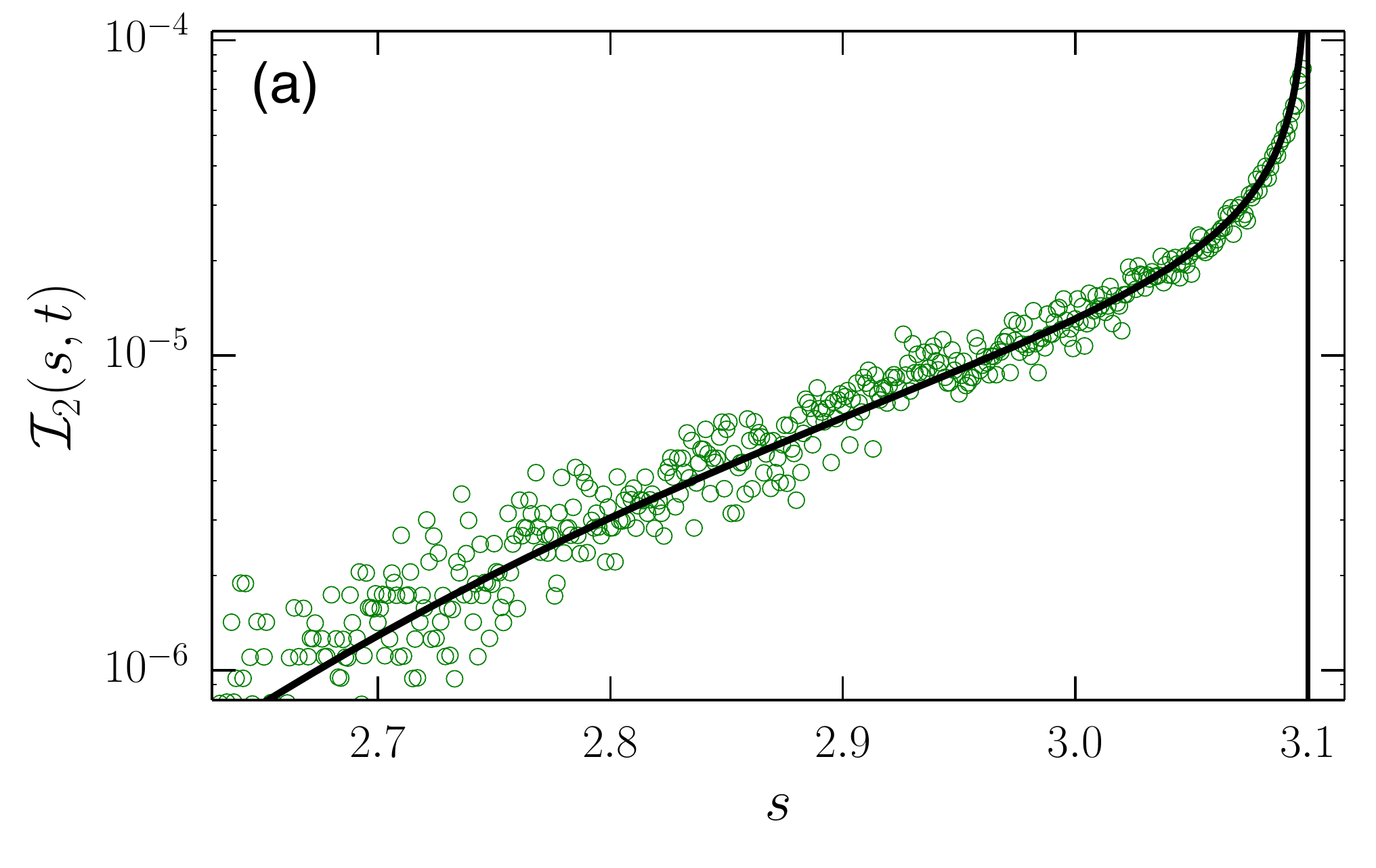}
\includegraphics[width = 0.45\textwidth]{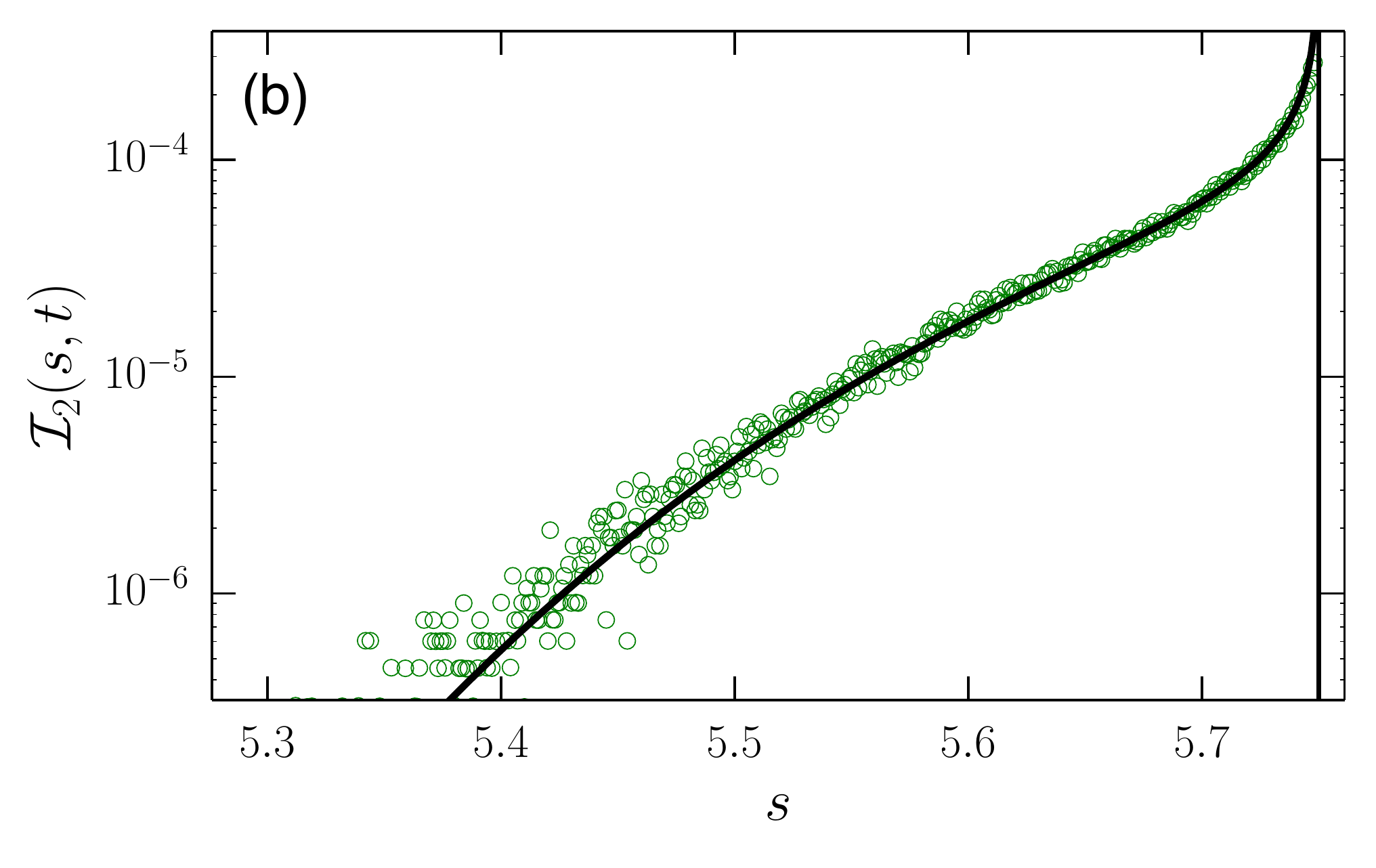}
\caption{Crossing probability for two disjunct intervals ${\cal I}_2(s,t)$ for the two cases shown in \fig \ref{fig:ensemble_sketch} for a fixed time $t$ indicated by the solid vertical black line at $t = 3.1$ (a) and $t=5.75$ (b) obtained from MC simulations (green circles) and \eq \eqref{eq:I2} (black line).
}
\label{fig:I2_direct}
\end{figure}

For ease of comparison, we provide separate panels for the first two peaks of the FPT distribution for both boundaries shown in \fig \ref{fig:ensemble_sketch}. For all cases ${\cal G}_1(t)$ lays on top of MC results and only starts to overestimate ${\cal F}(t)$ for times very close and larger than the local maximum of the FPT distribution. This is analogous to the Wiener--Rice expansion. For later times, it becomes more likely that the detected sign changes are not the first ones and hence do not constitute a FPT. We therefore count too many trajectories in ${\cal G}_1(t)$. To alleviate this problem, we need to subtract the probability for the trajectories that we falsely included in ${\cal G}_1(t)$. The leading correction term to ${\cal G}_1(t)$ accounts for all trajectories that have sign changes in $[t,t+\Delta]$ and $[s,s+\Delta]$ for $s < t$ \cite{verechtchaguina_2006}. Let $p_{+-+-}(s,t | x_0) =\mathbb P (X(s)>S(s),X(s+\Delta)<S(s+\Delta);X(t)>S(t),X(t+\Delta)<S(t+\Delta)|x_0)$ denote the probability for sign changes in two disjunct intervals, then
\begin{multline}
p_{+-+-}(s,t | x_0)=\\
\int_{S(t)}^\infty \!\!\!\ \! d x_3 \int_{-\infty}^{S(t)+\Delta} \!\!\!\! d x_4 f_2(x_3,x_4|x_0) p_{+-}(s|x_0,x_3,x_4)
\,,
\label{eq:p_pmpm}
\end{multline}
where $p_{+-}(s|x_0,x_3,x_4)$ is defined analogously to \eq \eqref{eq:pmp_td}, the only difference being that here we condition on three instead of one value of $X(t)$. By expanding \eq \eqref{eq:p_pmpm} to lowest order in $\Delta$ and using \eq \eqref{eq:p_pm_gauss}, we find
\begin{equation}
p_{+-+-}(s,t | x_0)=\sigma^2 \frac{\Delta}{\pi \tau} f_2(S(t),S(s)|x_0)\,.
\label{eq:p_pmpm-Gauss}
\end{equation}
Since $p_{+-+-}$ is a Gaussian, we can again integrate out the dependence on $x_0$ to obtain the probability ${\cal I}_2(s,t)=\int_0^\infty p_{+-+-}(s,t | x) p(x) d x$ for sign changes in two disjunct intervals $[s,s+\Delta]$ and $[t,t+\Delta]$ in closed form
\begin{equation}
{\cal I}_2(s,t)=\sigma^2 \frac{\Delta}{4 \pi^2 \tau} \frac{\exp[E(s,t)](1+\erf[F(s,t)]}{G(s,t))}\,.
\label{eq:I2}
\end{equation}
We provide details for $E(s,t)$, $F(s,t)$ and $G(s,t)$ in the Appendix. Note the structural similarity between \eqs \eqref{eq:I_1} and \eqref{eq:I2}. In \fig \ref{fig:I2_direct} we plot $\mathcal I_{2}(s,t)$ as a  function of $s$ for fixed values of $t$. There is almost no difference between the analytical expression and direct MC simulations, with the agreement ranging over three orders of magnitude in \fig \ref{fig:I2_direct}b. 
For a first-passage event at time $t$, second crossings can happen at any moment $s<t$. Therefore, we obtain the second order approximation ${\cal G}_2(t)$ of ${\cal F}(t)$ by summing ${\cal I}_2(s,t)$ over all possible values of $s$ that are commensurate with $\Delta$ and subtract this probability from ${\cal G}_1(t)$:
\begin{equation}
\mathcal G_{2}(t) = \mathcal I_{1}(t) - \sum_n \mathcal I_{2}(n \Delta,t) \,,
\label{eq:G_2}
\end{equation}
where we respect the strict inequality $n \Delta<t$. Figure \ref{fig:fpt_probabilities} shows that ${\cal G}_2(t)$ provides a significantly improved approximation of ${\cal F}(t)$ as compared with ${\cal G}_1(t)$. Note that ${\cal G}_2(t)$ generally captures the rising phase of the FPT distribution very well, even including the maximum in \fig \ref{fig:fpt_probabilities}b. This behaviour is consistent with other, although technically more limited, series expansions \cite{durbin_williams_1992,braun_matthews_thul_2015}. While ${\cal G}_1(t)$ tends to overestimate ${\cal F}(t)$, ${\cal G}_2(t)$ underestimates the true FPT statistics when it deviates at larger times. This is expected since we subtract too many trajectories in the computation of ${\cal G}_2(t)$ \cite{verechtchaguina_2006}, and can be remedied by including higher order terms in our expansion.

\begin{figure}
\centering
\includegraphics[width = 0.225\textwidth]{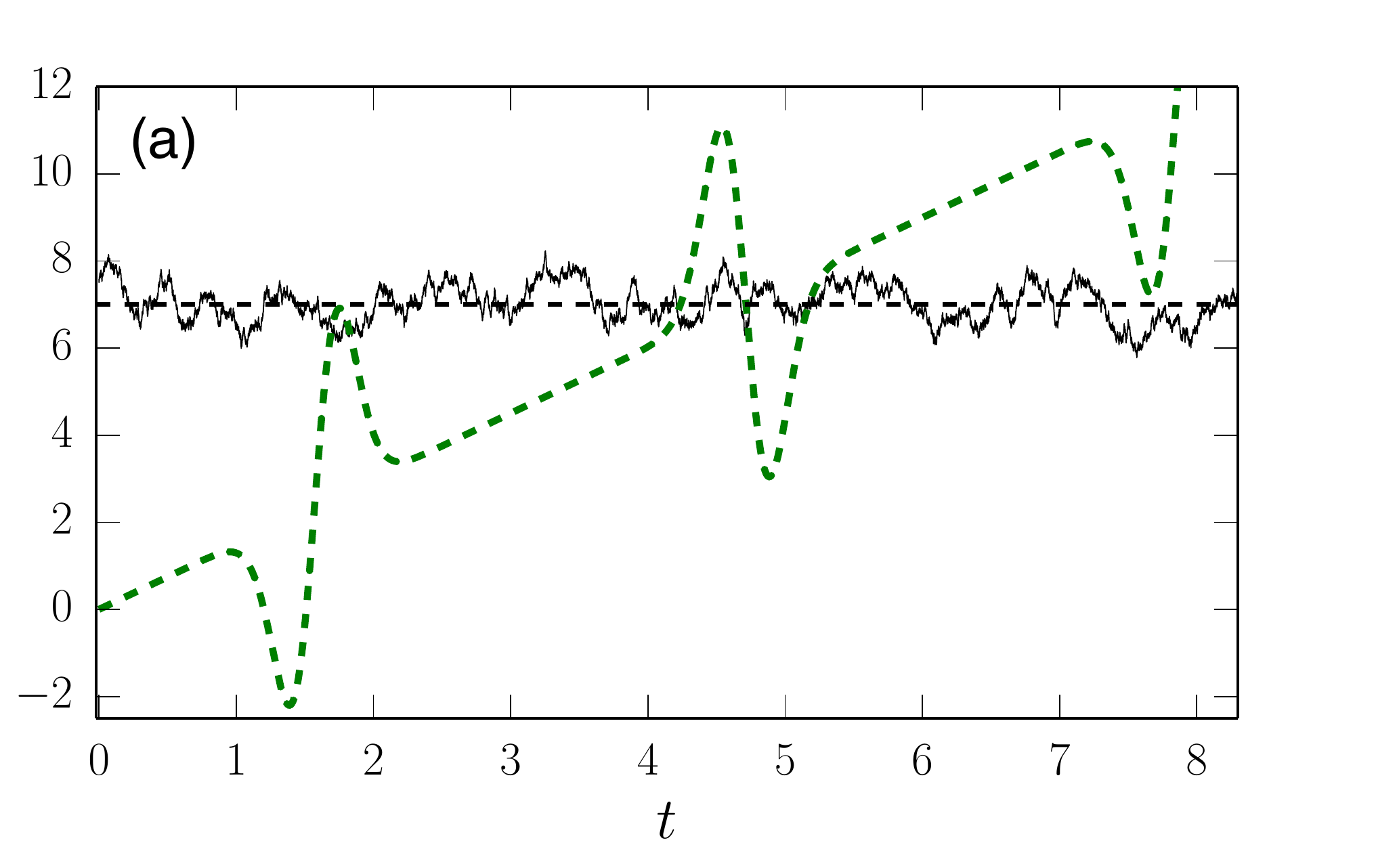}
\includegraphics[width = 0.225\textwidth]{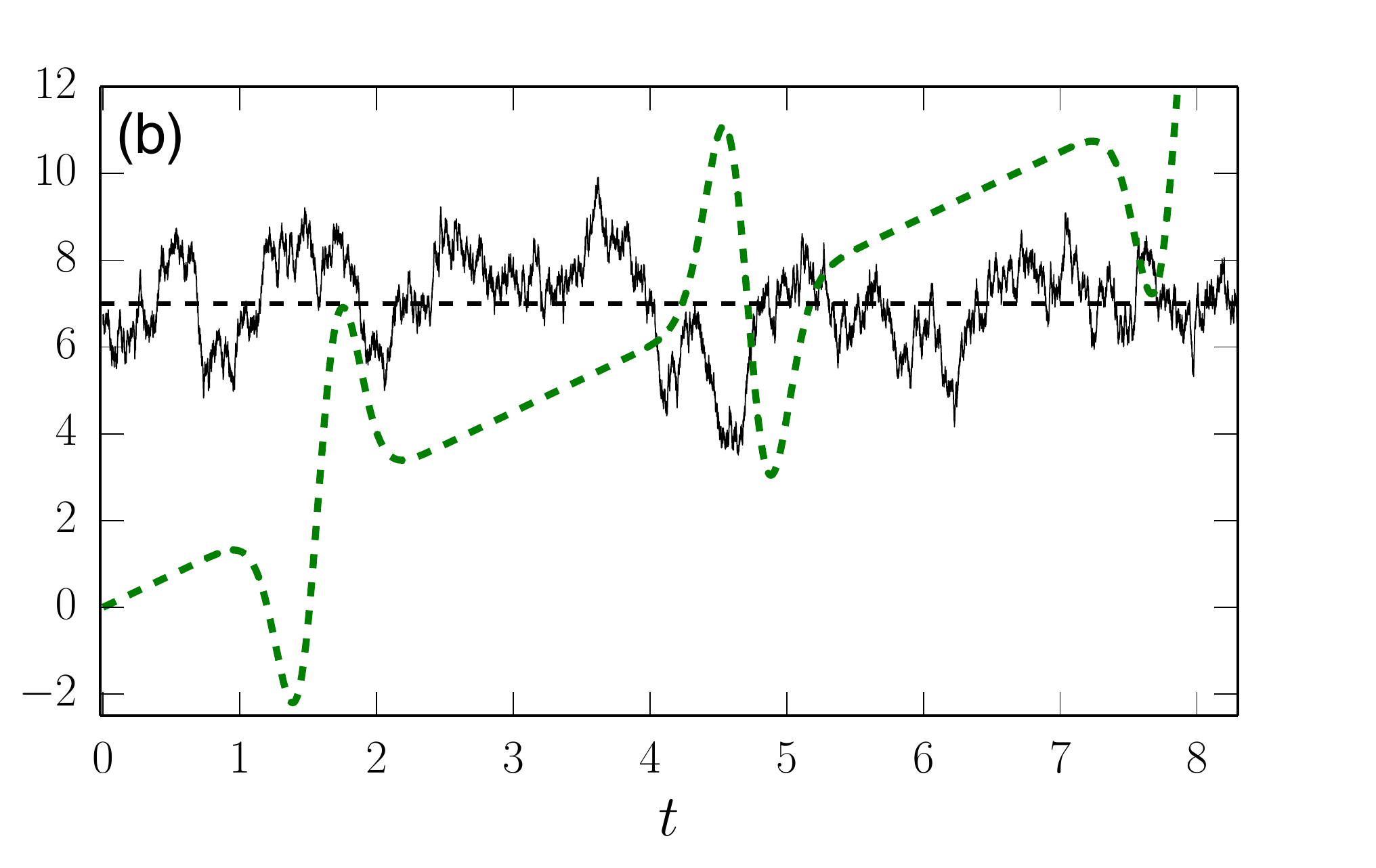}\\
\includegraphics[width = 0.225\textwidth]{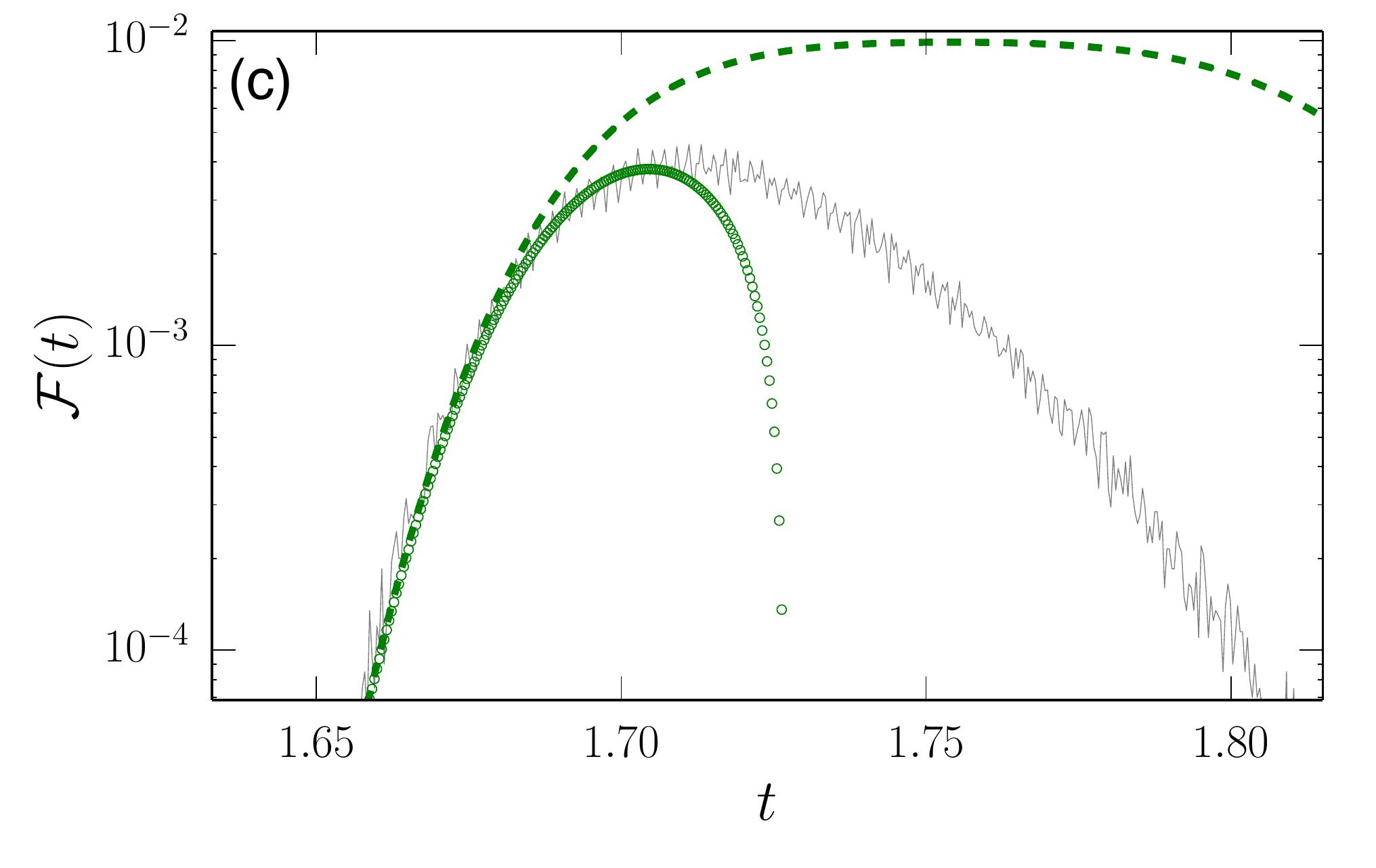}
\includegraphics[width = 0.225\textwidth]{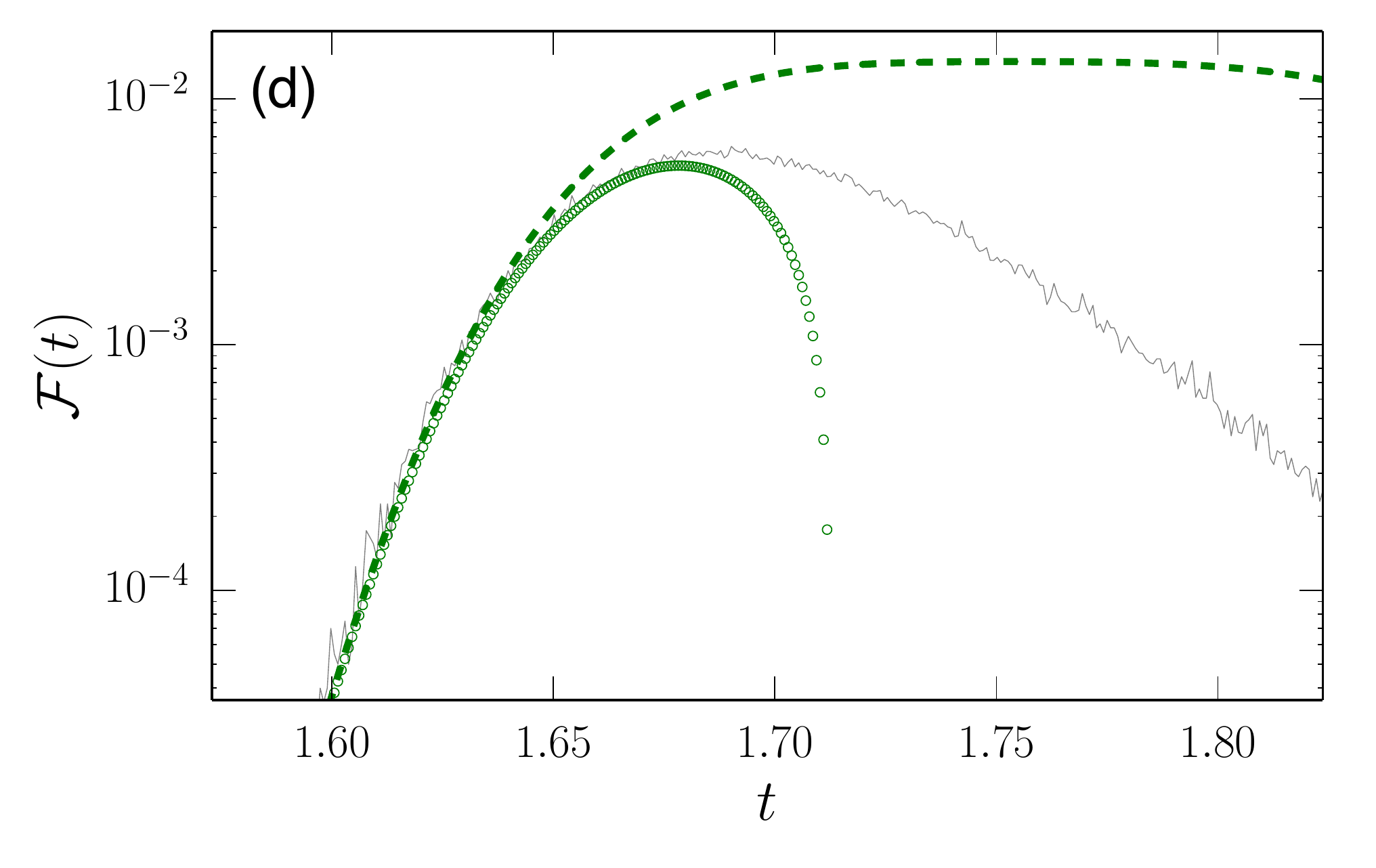}
\caption{(a,b) An OUP (solid black line) with mean $\vbar$ (dashed black line) crosses through a time dependent boundary $S(t)= -40\cos(t)\exp\left(-(4 \cos(t))^{2} \right) +  1.5 t$ (green dashed line). Parameter values are $\tau=0.2$, $\vbar=7$ and (a) $D=2.0$, (b) $D=10$. (c,d) FPT probabilities $\mathcal F(t)$ corresponding to (a) and (b) obtained from MC simulations (solid grey line), \eq \ref{eq:I_1} (dashed green line) and \eq \ref{eq:G_2} (green circles). 
}
\label{fig:large_noise}
\end{figure}

A particularly useful property of our approach is that it allows for explorations beyond the small noise limit. Figure \ref{fig:large_noise} shows results when the noise strength $D$ is increased by a factor of 5 above an already moderate noise level. As expected, the FPT distributions shift to smaller times for stronger noise, but the the agreement between numerics and analytics persists.

\begin{figure}[h!]
\centering
\includegraphics[width = 0.45\textwidth]{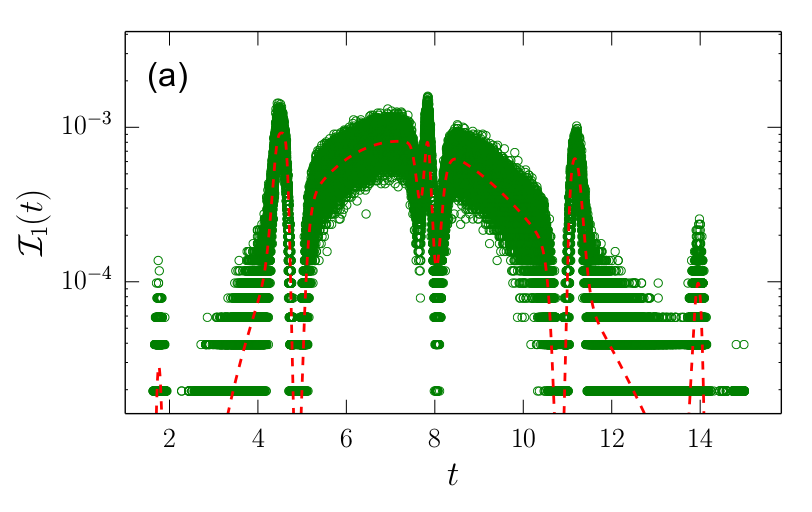}
\includegraphics[width = 0.45\textwidth]{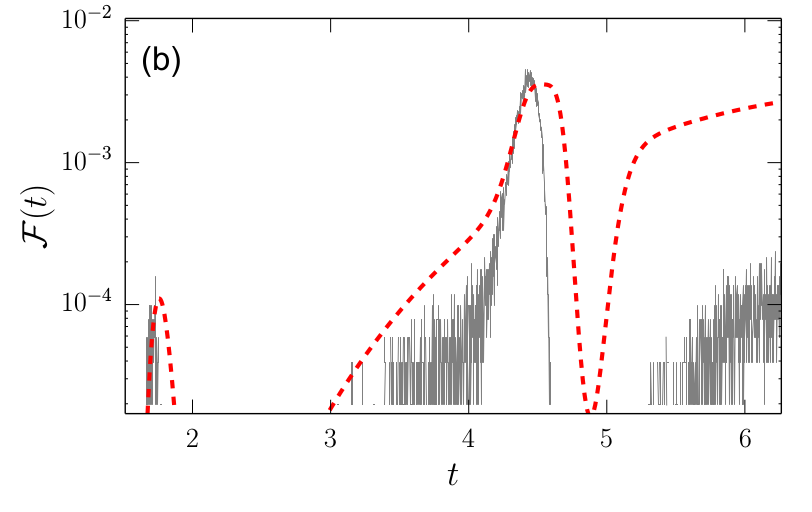}
\caption{(a) $\mathcal I_{1}(t)$ obtained from MC simulations (green circles) and \eq \ref{eq:I_1} (red dashed line). (b) FPT distribution obtained from MC simulations (grey lines) and  using for $p(x_0)=\delta(\vbar+2-x_0)$(red). The boundary is given by  $S(t)= -40\cos(t)\exp\left(-(4\cos(t))^{2} \right) + 1.5 t$, and the parameter values of the OUP are $\tau = 10.0$, $\vbar = 10.0$, $D=2.0$.}
\label{fig:initial_point_2}
\end{figure}

For the computation of ${\cal I}_1(t)$ and ${\cal I}_2(s,t)$ we integrated out the dependence on the initial values of the OUP. This could be done analytically since the OUP was initially Gaussian distributed. As the theory stands it can equally well deal with highly skewed initial distributions or sharp initial values as e.g. used in \cite{Schwalger:2015jl}. As an illustration of this point, we chose $x_0 \in [\vbar+2, \vbar+2.1]$, which corresponds to a strongly localised initial distribution. Figure \ref{fig:initial_point_2} shows results for ${\cal I}_1(t)$ and ${\cal F}(t)$. For the analytical results, we used $p(x_0)=\delta(\vbar+2-x_0)$ due to the narrow support of the distribution of $x_0$.
We observe reasonably good agreement between numerical and analytical results. Importantly, the theory is able to capture the small peak in the FPT distribution just before $t=2$. The main reason for the difference between the analytical and numerical results lies in the MC error. Indeed, the discrepancy between the analytical and numerical value for $\mathcal I_1(t)$ in \fig \ref{fig:initial_point_2}a at times $t \approx 2$ and $t \approx 14$ decreases with more MC realisations and hence is not a systematic error of the approach. 

\section{Discussion and conclusion}

In the present work, we have derived an analytical approach to FPT calculations for stationary non-differentiable Gaussian processes. Using the concept of sign changes, the results presented here overcome the often strong limitations of existing analytical techniques, such as convexity of the boundary, time scale separations or crossings being rare \cite{schindler_talkner_haenggi_2004, cramer_leadbetter_book, gs_review, durbin_williams_1992}, and allow us to construct high-fidelity approximations of the full FPT statistics, which are notoriously hard to compute. Our approach is based on the ideas of Wiener and Rice to appropriately count random trajectories and thus demonstrates that their seminal concept for differentiable Gaussian processes can be generalised to non-differentiable stochastic processes. Given that the computation of ${\cal G}_2(t)$ is entirely based on the evaluation and summation of Gaussian distributions, it is numerically fast and allows us to accurately describe events even when they occur with low probability, for which MC simulation are numerically expensive. While we illustrated our approach with the OUP, it works for any Gaussian process since equations \eqref{eq:p_pm_gauss} and \eqref{eq:p_pmpm} do not assume a specific form of the correlation coefficient $\rho(t)$. What the time-dependence of $\rho(t)$ determines is the scaling of ${\cal I}_1(t)$ and ${\cal I}_2(s,t)$ with $\Delta$. Interestingly, for the OUP with $\rho(t) \sim \exp(-|t|)$, we find ${\cal I}_1(t) \sim \sqrt{\Delta}$ when $t>\Delta/2$, a condition that is almost always satisfied due to the smallness of $\Delta$. By changing the correlation coefficient such that the stochastic process becomes differentiable, e.g. the widely used $\rho(t) \sim \exp(-t^2)$, we recover the original findings by Wiener and Rice  \cite{mcfadden_1967}. 

To illustrate the versatility of our approach, we chose examples to which, to the best of our knowledge, current analytical methods cannot be applied since the time dependent boundaries are neither convex nor concave. It might therefore be tempting to consider the stochastic process $Z(t)$ as introduced in \Sec \ref{sec:sign}, which is the difference between the stochastic process $X(t)$ and the time dependent boundary $S(t)$. Consequently, the FPT problem for $Z(t)$ involves a constant boundary at $0$. However, the stochastic differential equation for $Z(t)$ is explicitly time-dependent, which again renders the problem analytically intractable under general conditions \cite{touboul_faugeras_2007}. 

In the past, non-differentiable stochastic processes have often been dealt with by transforming them into differentiable versions via smoothing. This is usually done by either convolving the non-differentiable stochastic prcess with a kernel or by filtering the correlation function in frequency space \cite{Stratonovich:1967vm,gammaitoni_et_al_1991}. This requires firstly to find appropriate kernels or filters, and secondly to determine a bandwidth or cut-off frequency. This is not a trivial task, and often choices can only be justified \emph{a posteriori}. From a practical point of view, this is a severe limitation. In contrast, the approach presented here works directly with the original non-differentiable stochastic process and thus avoids any of these complications.

On theoretical grounds, we expect that the dominant contribution of ${\cal I}_1$ is in determining the rising phase of the peaks of the FPT distribution. This is well captured by the results shown in this study. Recently, the concept of few encounters has been introduced \cite{Godec:2016et}, and it was shown that the left part of the peaks of the FPT distribution is essential. Given the compact expression for ${\cal I}_1$, our results may prove useful in studying few encounters in more detail.

While we have focussed on stationary stochastic processes, an intriguing avenue for future research concerns the generalisation of our findings to non-stationary stochastic processes. The starting point is \eq \eqref{eq:p_pm_gauss}, but now both the mean and the standard deviation become time-dependent. Results analogous to \eqs \eqref{eq:I_1} and \eqref{eq:I2} would highlight the conceptual elegance of sign changing probabilities and moreover demonstrate that sign changing probabilities are a powerful and practical concept for investigating FPT and general level crossing \cite{azais_wschebor_book,feuerverger_hall_wood_1992} problems.

\begin{acknowledgments}
We would like to thank Paul Matthews for very insightful discussions and Dave Parkin for computational support. WB was partly supported by a Vice-Chancellor's Scholarship for Research Excellence at the University of Nottingham and NSERC Canada. 
\end{acknowledgments}

\appendix*

\section{Details for  \eq \eqref{eq:I2}}
We here provide details for the sign changing probability in two disjunct intervals, \eq \eqref{eq:I2}. The functions $E(s,t)$ and $F(s,t)$  are given by $E(s,t)=A(s,t)+B^2(s,t)/\Gamma(s,t)$ and $F(s,t)=B(s,t)/\sqrt{\Gamma(s,t)}$ with
\begin{align*}
A(s,t) & = -\frac{\vbar^{2}}{2 \sigma^{2}} - \frac{\alpha(s,t)}{2(1-\widetilde{\rho}^{2})}\,,\\
B(s,t) &= \frac{\vbar}{\sigma^{2}} - \frac{\beta(s,t)}{2(1-\widetilde{\rho}^{2})}\,,\\
\Gamma(s,t) &= 2 \left[\frac{1}{\sigma^{2}} + \frac{\gamma(s,t)}{1-\widetilde{\rho}^{2}} \right]\,, 
\end{align*}
and
\begin{align*}
\alpha(s,t) &= \frac{S_{s}-\vbar(1-\rho_{s}^{2})}{\sigma_{s}^{2}} + \frac{S_{t}-\vbar(1-\rho_{s}^{2})}{\sigma_{t}^{2}}
-\frac{2\tilde{\rho}}{\sigma_{s}\sigma_{t}}
[S_{s}S_{t}\\
+S_{s}&\vbar(\rho_{t}-1)+S_{t}\vbar(\rho_{s}-1)+\vbar^{2}(1-\rho_{t}-\rho_{s}+\rho_{s}\rho_{t}) ] \,,\\
\beta(s,t) &= \frac{2(\vbar(1-\rho_{s})\rho_{s}-S_{s}\rho_{s})}{\sigma_{s}^{2}} + \frac{2(\vbar(1-\rho_{t})\rho_{t}-S_{t}\rho_{t})}{\sigma_{t}^{2}}\\
&-\frac{2\tilde{\rho}}{\sigma_{s}\sigma_{t}}\left[-S_{s}\rho_{t}-S_{t}\rho_{s}+\vbar\rho_{t}+\vbar\rho_{s}-2\vbar\rho_{s}\rho_{t}\right] \,,\\
\gamma(s,t)&=\frac{\rho_{s}^{2}}{\sigma_{s}^{2}} + \frac{\rho_{t}^{2}}{\sigma_{t}^{2}} -\frac{2\tilde{\rho}}{\sigma_{s}\sigma_{t}}\rho_{s}\rho_{t} \,,
\end{align*}
while $G(s,t)$ reads as
\begin{align*}
G(s,t)&=\sqrt{(1-\tilde{\rho}^{2})\sigma_{s}^{2}\sigma_{t}^{2}+\sigma^{2}\left(\rho_{s}^{2}\sigma_{t}^{2}+\rho_{t}^{2}\sigma_{s}^{2}-2\tilde{\rho}\sigma_{s}\rho_{s}\sigma_{t}\rho_{t}\right)}\,,\\
\widetilde{\rho}(s,t) &= \frac{\rho(s-t) - \rho(s) \rho(t)}{\sqrt{1-\rho(t)^2}\sqrt{1-\rho(s)^2}}\,.
\end{align*}
For notational convenience, subscripts refer to time arguments, e.g. $\sigma_{t} = \sigma(t)=\sigma \sqrt{1-\rho^{2}(t)}$ and $S_t=S(t)$.

\bibliography{literature_paper}

\end{document}